\documentclass{nature}

\usepackage{graphicx}

\usepackage{color}
\usepackage{epsfig}

\title{Near-room-temperature giant topological Hall effect in \\antiferromagnetic kagome metal YMn$_{6}$Sn$_{6}$}

\author{Qi Wang,$^{1}$ Qiangwei Yin,$^{1}$ Satoru Fujitsu,$^{2}$ Hideo Hosono,$^{2,3}$ and Hechang Lei$^{1,*}$}

\begin{document}

\maketitle

\begin{affiliations}
\item Department of Physics and Beijing Key Laboratory of Opto-electronic Functional Materials $\&$ Micro-nano Devices, Renmin University of China, Beijing 100872, China
\item Materials Research Center for Element Strategy, Tokyo Institute of Technology, Yokohama 226-8503, Japan
\item Laboratory for Materials and Structures, Tokyo Institute of Technology, Yokohama 226-8503, Japan
\end{affiliations}

\leftline{$^*$Corresponding author: hlei@ruc.edu.cn}\vspace*{1cm}

\begin{abstract}
The kagome lattice, consisting of interconnected triangles and hexagons uniquely, is an excellent model system for study frustrated magnetism, electronic correlation and topological electronic structure.
After an intensive investigation on frustrated magnetism in insulating magnetic kagome lattices, the interplay between charge and spin degrees of freedom via spin-orbital coupling in metallic systems (kagome metals) has become an attractive topic recently.
Here, we study centrosymmetric YMn$_{6}$Sn$_{6}$ with Mn kagome lattice. We discover that it exhibits giant topological Hall effect near room temperature, ascribed to the field-induced non-collinear spin texture, possibly a skyrmion lattice (SkL) state.
Combined with the large intrinsic anomalous Hall effect, YMn$_{6}$Sn$_{6}$ shows a synergic effect of real- and momentum-space Berry phase on physical properties.
Since the features of tunable magnetic interaction and flexible structure in this large homologous series, it provides a novel platform for understanding the influence of electronic correlations on topological quantum states in both real and momentum spaces.
\end{abstract}

Berry phase effects on electronic properties have become one of central topics in condensed matter physics.\cite{Berry,XiaoD}
Meanwhile, topology of electronic structure, through Berry's phase, introduces a novel classification method of matter and a series of non-trivial topological matters have been proposed in theory and verified in experiment, such as topological insulators, Dirac and Weyl semimetals etc.\cite{Hasan1,QiXL1,WanX,WangZ,LiuZK1,WengH,XuSY,LvBQ}
For the last several decades, it is also realized that a vast number of exotic phenomena deeply root in the momentum-space Berry phase and closely related Berry curvature, such as quantum Hall effect,\cite{Thouless1} intrinsic (quantum) anomalous Hall effect (AHE) and magnetic monopole in $k$-space,\cite{Jungwirth,Onoda1,Haldane,Nagaosa1,YuR1,ZhangCZ,Fang} (quantum) anomalous Hall effect,\cite{Murakami,Sinova,Kane1,Konig} chiral anomaly and Fermi arcs at the surface state in topological semimetals.\cite{SonDT,HuangX,WanX}
On the other hand, real-space Berry phase originating from non-collinear spin texture or magnetic topological excitations like skyrmions with non-zero scalar spin chirality $\textbf{S}_{i}\cdot \textbf{S}_{j}\times \textbf{S}_{k}$ (where $\textbf{S}_{i,j,k}$ are spins) has also manifest its influence on electronic properties. The emergent electromagnetic field related to the real-space Berry phase plays the same role as a real magnetic field in the charge dynamics, thus it can also produce a Hall effect, called topological Hall effect (THE).\cite{Binz,YeJ,Taguchi,Neubauer2}

Insulating magnetic kagome lattices have been studied intensively for a long time because of the potential existence of frustration-induced quantum spin liquid state.\cite{ZhouY,HanTH} More importantly, theory predicts that magnetic kagome lattice can also host non-trivial topological states with unusual physical properties,\cite{XuG1,Mazin,TangE,ChenH,Kubler} but these features are not observed in experiment until very recently.\cite{Nakatsuji,Nayak,Kuroda,YeL,YinJX,LinZ,LiuE,WangQ1,WangQ2}
a spectrum of exotic phenomena has been observed in metallic compounds with kagome lattice of 3$d$ transition metals, named kagome metals. For example, there is a large AHE originating from the momentum-space Berry phase of massive Dirac or magnetic Weyl nodes in Fe$_{3}$Sn$_{2}$, Mn$_{3}$X (X = Ge, Sn) and Co$_{3}$Sn$_{2}$S$_{2}$,\cite{WangQ1,Nakatsuji,Nayak,Kuroda,YeL,LiuE,WangQ2} and Fe$_{3}$Sn$_{2}$ exhibits flat-band electronic structure and giant magnetization-driven electronic nematicity.\cite{LinZ,YinJX}
In contrast to such rich interplay between magnetism and quantum electronic structure in kagome metals, the non-collinear spin texture (real-space Berry phase) induced electromagnetic responses of conduction electrons are still not well understood.\cite{Nakatsuji,HouZ}

In this work, we study the magnetoelectric responses in antiferromagnetic kagome metal YMn$_{6}$Sn$_{6}$, composed of Mn kagome lattice in the centrosymmetric hexagonal structure. It shows a giant THE near room temperature after the spin-flop transition. Moreover, at high field region, the large intrinsic AHE has also been observed when the spins are fully polarised along the $a$ axis. Both effects could originate from real- and momentum-space Berry phases, respectively.
These unique features of YMn$_{6}$Sn$_{6}$ suggest that the interplay of tunable real-space spin texture, non-trivial band structure and electronic correlations would lead to abundant exotic behaviors in magnetic Kagome metals.


\section*{Structure and metallic antiferromagnetism in YMn$_{6}$Sn$_{6}$}

YMn$_{6}$Sn$_{6}$ has the hexagonal MgFe$_{6}$Ge$_{6}$-type structure with space group of $P6/mmm$ (No. 191) and hundred of compounds belong to this structural family. The refined lattice parameters are $a=$ 5.5362(2) \AA\ and $c=$ 9.0146(3) \AA\ (Supplementary Fig. 1).\cite{Malaman} As shown in Fig. 1a, there are three kinds of Sn sites and the crystal structure of YMn$_{6}$Sn$_{6}$ is composed of Y-Sn3 layers and Mn-Sn1-Sn2-Sn1-Mn slabs stacking along the $c$ axis alternatively. For the Y-Sn3 layer, the Sn3 atoms form graphene-like hexagon planes and the Y atoms occupy (lie in) the centers of the hexagons in Sn3 layers. For the Mn-Sn1-Sn2-Sn1-Mn slab, the Mn atoms form two kagome layers with $d_{\rm Mn-Mn}$ ($\sim$ 2.76810(9) \AA) and the Sn2 atoms from a graphene-like hexagon layer in between, which is same as the Sn3 atoms. Moreover, like the Y atoms in the Y-Sn3 layer, the Sn1 atoms also locate at the centers of the hexagons of Mn layers but they are far below or above those Mn layers. This is distinct from Co$_{3}$Sn$_{2}$S$_{2}$ and Fe$_{3}$Sn$_{2}$, where the Sn atoms (nearly) lie in the Co or Fe Kagome layers. Actually, the structure of YMn$_{6}$Sn$_{6}$ is closely related to the CoSn-B35 type of structure, where the sites of Y atoms are empty and the Sn1 is exactly in the Co Kagome layer. Thus, the occupancy of Y atoms should expel the Sn1 atoms away from the Mn Kagome layer. It leads to a very short bond distance between two Sn1 atoms ($d_{\rm Sn1-Sn1}\sim$ 2.87024(9) \AA), much shorter than that in CoSn ($d_{\rm Sn1-Snx1}\sim$ 4.25 \AA).\cite{Larsson} In contrast, the filling of Y atoms has minor influence on the $d_{\rm Sn3-Sn3}$ ($\sim$ 3.1963(1) \AA), which is same as the $d_{\rm Sn2-Sn2}$ surprisedly and comparable to that in CoSn ($d_{\rm Sn1-Sn1}\sim$ 3.05 \AA).\cite{Larsson} Because of such unique structural feature, the magnetic properties of RMn$_{6}$Sn$_{6}$ are very sensitive not only to the $d_{\rm Mn-Mn}$ in the Mn Kagome layer but also to the nature of R elements. The moments of Mn atoms lie in the $ab$-plane and the interplane Mn moments through the Mn-Sn1-Sn2-Sn1-Mn slab are always parallel (ferromagnetic), but those occurring through the Mn-(R-Sn)-Mn slab rotate with a non-constant angle (antiferromagnetic), depending on the R elements in detail.\cite{Venturini}
Consistent with the layered structure with strong interlayer interaction, YMn$_{6}$Sn$_{6}$ single crystal exhibits a thick plate-like shape and the $c$ axis is perpendicular to the crystal surface (Supplementary Fig. 1).

Fig. 1b shows the temperature dependence of magnetization $M(T)$ at $B=$ 0.5 T for the field parallel to two different crystallographic directions. It can be seen that for $B\Vert (100)$, the $M(T)$ curve shows two sequential transitions at about $T_{N}\sim$ 359 K and $T_{t}\sim$ 326 K, close to previous results.\cite{Matsuo,Uhlirova} The former one is corresponding to the paramagnetism-antiferromagnetism transition and the latter one accompanying with the sudden drop of the $M(T)$ curve could be ascribed to the collinear to helical AFM transition.\cite{Matsuo}
In contrast, the $M(T)$ curve for $B\Vert (001)$ is rather flat below $T_{N}\sim$ 359 K, a typical behavior when the easy axis or plane of magnetization is in the $ab$ plane. All of zero-field cooling (ZFC) and field-cooling (FC) $M(T)$ curves overlap each other very well, clearly indicating the absence of magnetic glassy state in YMn$_{6}$Sn$_{6}$.
Isothermal magnetization $M(B)$ at $T=$ 3 K are shown in Fig. 1c. The absence of hysteresis for all of $M(B)$ curves confirms the AFM state below $T_{N}$. For $B\Vert (001)$, the $M(B)$ increases smoothly because of gradual bending of the moments towards the $c$ axis.
But the $M(B)$ curve for $B\Vert (100)$ shows a striking jumps at $B_{\rm sf}\sim$ 2.5 T, derived from the d$M(B)$/d$B$ curve (inset of Fig. 1c), leading to larger $M$ than that for $B\Vert (001)$. It is reminiscent of a typical spin-flop process, implying the weak magnetocrystalline anisotropy in YMn$_{6}$Sn$_{6}$.
The slope of $M(B)$ curve starts to decrease when $B>B_{\rm sat,s}$ ($\sim$ 10.4 T) and finally the fully polarised ferromagnetic (PFM) state is reached at $B_{\rm sat,e}\sim$ 13.2 T. The saturated moment for Mn atom is about 2.26 $\mu_{B}$/Mn for $B\Vert (100)$.
It should be noted that there is another small kink located at $B_{\rm t}\sim$ 7.5 T, which has also been observed in previous studies.\cite{Matsuo,Uhlirova} It may reflect more than one AFM interactions in YMn$_{6}$Sn$_{6}$, closely related to the structural feature along the $c$ axis.
The $M(T)$ curves at 5 T and 14 T for both field directions are also consistent with above $M(B)$ results (Supplementary Fig. 2).

Zero-field longitudinal resistivity $\rho_{xx}(T)$ and $\rho_{zz}(T)$ display metallic behaviors between 3 and 400 K (Fig. 1d). There is an increase of slope of resistivity curves below $T_{N}$, corresponding to the suppression of spin disorder scattering.
The ratio of $\rho_{zz}(T)/\rho_{xx}(T)$ in the whole measuring temperature range is about 0.7 - 0.9 (inset of Fig. 1d), suggesting that the $c$-axial conductivity is even larger than that in the $ab$ plane and there is a weak anisotropy in this compound, i.e., YMn$_{6}$Sn$_{6}$ should be regarded as a three-dimensional (3D) rather than two-dimensional (2D) metallic antiferromagnet.

\section*{AHE and THE of YMn$_{6}$Sn$_{6}$}

The Hall resistivity $\rho_{yx}$ of YMn$_{6}$Sn$_{6}$ increases with field at low-field region and then becomes saturated when increasing field further (Fig. 2a). The saturation field shifts to higher field with lowering $T$ and when $T\leq$ 80 K, the saturation region can not be observed up to 14 T.
At first glance, the behavior of $\rho_{yx}$ closely resembles that of $M(B)$ at corresponding temperature (Fig. 2c), but there is a subtle difference between them, especially at low temperatures where the $M(B)$ curves are convex and the $\rho_{yx}$ curves are slightly concave (Supplementary Fig. 3).
For longitudinal resistivity $\rho_{xx}(B)$, it exhibits negative magnetoresistance (MR) at hight temperatures and when temperature decreases, the positive MR appears at low-field region but starts to decrease at high-field region (Fig. 2b).
On the other hand, at high temperatures the $\rho_{yz}$ of sample fabricated by using the focus ion beam (FIB) technique (Supplementary Fig. 4) also increases quickly with field, but it decreases slightly at high field region (Fig. 2c). With decreasing temperature, the $\rho_{yz}$ curve shows a sharp jump at the spin-flop field (Fig. 2f). One of the most striking features of $\rho_{yz}$ curve is the obvious hump between $B_{\rm sf}$ and $B_{\rm sat,e}$ and such behavior is distinctly different from the nearly linear increase and gradual saturation behaviors of $M(B)$ with field (Supplementary Fig. 3). This feature becomes weaker when lowering $T$ and another interesting feature that $\rho_{yz}$ dips around 10 T appears, which has been observed in previous study.\cite{Uhlirova}
The $\rho_{zz}(B)$ shows similar field dependence to $\rho_{xx}(B)$ in general at the whole temperature and field ranges (Fig. 2e), but it exhibits more obvious kinks near those characteristic fields, such as $B_{\rm sf}$ and $B_{\rm sat,s}$ etc. It implies that the changes of magnetic configurations have remarkable influence on longitudinal resistivity, especially for $\rho_{zz}$.

In a magnetic system, the total Hall resistivity usually can be described as the sum of three contributions:\cite{Nagaosa1,Nagaosa2} $\rho_{\rm H} = \rho_{\rm H}^{\rm N} + \rho_{\rm H}^{\rm A} + \rho_{\rm H}^{\rm T} = R_{0}B + S_{\rm H}\rho^{2}M + \rho_{\rm H}^{\rm T}$, where $\rho_{\rm H}^{\rm N}$ is the normal Hall resistivity due to the Lorentz force and $R_{0}$ is the ordinary Hall coefficient. $\rho_{\rm H}^{\rm A}$ is the anomalous Hall resistivity and because the intrinsic anomalous Hall conductivity (AHC, $\sigma_{\rm H}^{\rm A}\sim \rho_{\rm H}^{\rm A}/\rho^{2}$) is proportional to $M$ linearly, $S_{\rm H}$ should be a constant for intrinsic AHE.\cite{Nagaosa1,ZengC}
The last term $\rho_{\rm H}^{\rm T}$ represents the topological Hall resistivity, usually originating from non-collinear spin texture with non-zero scalar spin chirality.
The curves of $\rho_{\rm H}/B$ vs. $\rho^{2}M$ show perfectly linear behaviors at high-field region for both $\rho_{yx}$ and $\rho_{yz}$ when $T>$ 80 K (Supplementary Fig. 3). It undoubtedly indicates that the AHE is dominant when the THE vanishes at the fully polarised FM state. The high-field linear region can not be observed at $T\leq$ 80 K (Supplementary Fig. 3) because the saturation region is beyond 14 T for $\rho_{yx}$ or the emergence of upturn behavior for $\rho_{yz}$.
The determined $R_{0}$ and $S_{\rm H}$ as the intercept and slope of the scaling curves above the critical field $B_{\rm sat,e}$ are shown in Figs. 3a and 3d, respectively. The $R_{0}$ derived from $\rho_{yx}$ is positive and almost independent of temperature in contrast to that obtained from $\rho_{yz}$ which shows negative values at high temperatures and a strong temperature dependence with a sign change at $T\sim$ 120 K.
The calculated apparent carrier density $n_{a}$ ($=1/|e|R_{0}$, $e$ is the elementary charge) (Supplementary Fig. 5) at 320 K is about 1.22 and -0.86$\times$10$^{22}$ cm$^{-3}$, i.e., about 2.90 and -2.05 per formula unit of YMn$_{6}$Sn$_{6}$, for $B\Vert (001)$ and $B\Vert (100)$. The opposite sign of $R_{0}$ (or $n_{a}$) implies that magnetic field may have a significant effect on the electronic structure at PFM state.
Both $S_{\rm H}$ obtained from $\rho_{yx}$ and $\rho_{yz}$ are positive. The latter is about 6 times larger than the former and its absolute value is comparable with those in typical FM materials and known kagome metals, such as Fe (0.06 V$^{-1}$),\cite{Dheer} MnSi (-0.19 V$^{-1}$),\cite{Neubauer} Mn$_{3}$Sn (0.07 - 0.24 V$^{-1}$),\cite{Nakatsuji} Fe$_{3}$Sn$_{2}$ (0.04 - 0.09 V$^{-1}$),\cite{WangQ1} and Co$_{3}$Sn$_{2}$S$_{2}$ (0.05 V$^{-1}$).\cite{WangQ2}
In addition, the $\sigma_{\rm H}^{\rm A}$ 
at 14 T for both field directions shows a weak dependence on longitudinal conductivity $\sigma$ ($\sim 1/\rho$) and the scaling index $\alpha$ in $\sigma_{\rm H}^{\rm A}\propto\sigma^{\alpha}$ equals 0.35(6) and 0.4(1) for $\sigma_{xy}^{\rm A}$ and $\sigma_{zy}^{\rm A}$, respectively (Supplementary Fig. 6). Thus, above results indicate that the AHE at high-field region is indeed intrinsic with relative large values of $\sigma_{xy}^{\rm A}$ ($\sim$ 45 S cm$^{-1}$) and $\sigma_{zy}^{\rm A}$ ($\sim$ 300 S cm$^{-1}$).

The $\rho_{\rm H}^{\rm T}$ is estimated after the subtraction of $\rho_{\rm H}^{\rm N}$ and $\rho_{\rm H}^{\rm A}$ from $\rho_{\rm H}$ (Supplementary Fig. 3). As shown in Figs. 3c and 3d, at high temperature, both $\rho_{yx}^{\rm T}$ and $\rho_{yz}^{\rm T}$ are small. Their intensities increase with decreasing temperature at first, reaching the maximum value of $\rho_{yx}^{\rm T,max}=$ -0.28 $\mu\Omega$ cm and $\rho_{yz}^{\rm T,max}=$ 2.0 $\mu\Omega$ cm at about 220 K and 240 K, and then decrease quickly.
At lower temperature, the $\rho_{yx}^{\rm T,max}$ shifts to higher field. In contrast, the $\rho_{yz}^{\rm T,max}$ becomes negative when $T\leq$ 80 K, located at $B\sim$ 10 T.
The combined phase diagrams of $\rho_{\rm H}^{\rm T}$ and magnetic transitions (Figs. 3e and 3f) provide a clear relation between field-induced magnetic structures and emergent electrodynamic responses.
For $\rho_{yx}^{\rm T}$, with increasing field, the spin direction gains a $c$-axial component in the helical AFM (H-AFM) state and this might lead to non-collinear and/or non-coplanar spin configuration, contributing to the THE. When $B>B_{\rm sta,s}$, the beginning of spin polarization destroys above spin texture, resulting in the decrease of THE.
On the other hand, below $B_{\rm sf}$, the $\rho_{yz}^{\rm T}$ is also small, same as $\rho_{yx}^{\rm T}$ with similar reason. When $B_{\rm sf}<B<B_{\rm sat,e}$ at high temperature, the spin-flop process generates a topologically nontrivial magnetic structure (AFM-1 state) with large scalar spin chirality,\cite{Uhlirova} and thus there is huge response of THE. When $T<$ 80 K, there may be somewhat changes in the AFM-1 state, such as increase of modulation period or the disappearance of skyrmion lattice phase which only survives at the small $T-B$ region just below $T_{N}$ with the assistance of thermal fluctuations,\cite{Muhlbauer} and the THE diminishes at same region as well. But the higher field ($B_{\rm t}<B<B_{\rm sat,e}$) at lower temperature seems to induce another magnetic structure (AFM-2 state) with non-zero scalar spin chirality, and correspondingly the THE becomes pronounced again with the opposite sign. Finally, once $B$ is beyond $B_{\rm sat,e}$, the PFM state can not host a topologically non-trivial spin structure, therefore the THE disappears.

\section*{Discussion}

In the following, we discuss the THE in the framework of the skyrmion-induced scalar spin chirality. First, the value of $\rho_{yz}^{\rm T,max}$ (2.0 $\mu\Omega$ cm) is much larger than those of materials hosting SkL, such as B20-type MnSi (0.04 $\mu\Omega$ cm under high pressure) and MnGe (0.16 $\mu\Omega$ cm),\cite{} and Heusler compounds Mn$_{2}$RhSn (0.05 $\mu\Omega$ cm) and Mn$_{2}$PtSn films (0.57 $\mu\Omega$ cm).\cite{Rana,LiY}
In the skyrmion systems, the THE can be expresses as $\rho_{H}^{\rm T}=PR_{0}B_{\rm eff}^{z}$, where $P$ is the local spin polarisation of the conduction electrons and $B_{\rm eff}^{z}$ is the effective magnetic field.\cite{Neubauer2}
The value of $P$ can be estimated by using $P=\mu_{\rm spo}/\mu_{\rm sat}$, where $\mu_{\rm spo}$ is the ordered magnetic moment in the Skyrmion phase (AFM-1 state), and $\mu_{\rm sat}$ is the saturated moment deduced from the Curie-Weiss law in the paramagnetic state above $T_{N}$.\cite{Neubauer2} With $\mu_{\rm spo}=$ 1.11 $\mu_{\rm B}$/Mn (5 T and 240 K for $B\Vert (100)$) and $\mu_{\rm sat}=$ 3.61 $\mu_{\rm B}$/Mn,\cite{Venturini2} we obtain $P=$ 0.307. Using the $R_{0}=$ -0.00157 cm$^{3}$/C and $\rho_{yz}^{\rm T,max}=$ 2.0 $\mu\Omega$ cm at 240 K, the estimated $B_{\rm eff}^{z}$ induced by SkL is about -41.5 T. Assuming the triangular SkL, it has $B_{\rm eff}^{z}=-\frac{\sqrt{3}\phi_{0}}{2\lambda^{2}}$, where $\phi_{0}=h/|e|$ is the flux quantum for a single electron with the Planck constant $h$, and $\lambda$ is the helical period.\cite{Nagaosa2} The calculated $\lambda$ is $\sim$ 6.6 nm, smaller than most of B20-type helimagnets.\cite{Nagaosa2} Thus, the larger THE in YMn$_{6}$Sn$_{6}$ than those in non-centrosymmetric materials hosting SkL could partially originate from one or two orders of magnitude higher $R_{0}$, remarkably larger $P$ and shorter wavelength of the spin modulation. It has to be noted that the $P$ might be overestimated because it usually depends on a complicated Fermi-surface average.\cite{Ritz}
Because the $\rho_{H}^{\rm T}$ is considered to depend on the $P$ that is very sensitive to the band structure, the sign change of $\rho_{yz}^{\rm T}$ at high field and low temperature might closely related to the evolution of magnetic structure which change the band structure subtly.\cite{LiYF}
Another possible reason leading to the sign inverse of $\rho_{yz}^{\rm T}$ is the change of chirality of skyrmions. It usually can not happen for the SkL induced by the Dzyaloshinskii-Moriya (DM) interaction in non-centrosymmetric materials because the chirality is determined by the sign of DM interaction, but in the case of the SkL originating from the frustrated exchange interactions, the skyrmion and anti-skyrmion with opposite chiralities are degenerate.\cite{Nagaosa2}

Second, since the structure of YMn$_{6}$Sn$_{6}$ is centrosymmetric, it should exclude the origin of skyrmion state from the DM interaction.
Moreover, the size of the skyrmions in YMn$_{6}$Sn$_{6}$ is rather small, also inconsistent with that induced by the long-ranged magnetic dipolar interactions (100 nm to
1 $\mu$m).\cite{Nagaosa2} On the other hand, frustrated exchange interactions could lead to the small size of skyrmion size.\cite{Okubo} In YMn$_{6}$Sn$_{6}$, the magnetic interaction along the $c$-axis is frustrated when considering the interactions through the Mn-Sn1-Sn2-Sn1-Mn path, the Mn-(R-Sn)-Mn path and the next-nearest-neighbour Mn-Sn1-Sn2-Sn1-Mn-(R-Sn)-Mn path.\cite{Venturini} The partial helical AFM reflects this feature clearly.\cite{Venturini} In addition, when $B>B_{\rm sf}$, the flopped spins sitting on the kagome lattice could be also frustrated in the $ab$-plane. All of these frustrations may cause rich multiple-$Q$ structures and the strong coupling between itinerant electrons and these unconventional magnetic structures of local Mn moments results in the giant emergent electromagnetic responses.
Similar behaviour is also observed in centrosymmetric Gd$_{2}$PdSi$_{3}$ very recently.\cite{Kurumaji} In there, the huge THE is induced by the skyrmion state stabilized by the frustration of Gd 4f moments on the triangular lattice.

Third, besides the giant THE, YMn$_{6}$Sn$_{6}$ also exhibits large field-direction-dependent AHE. Pervious studies on kagome metals indicate that they are topological semimetals harboring massive Dirac or Weyl fermions that root in the kagome lattices composed of magnetic elements.\cite{YeL,LiuE,WangQ1,WangQ2} YMn$_{6}$Sn$_{6}$ with Mn kagome lattice may also exhibit similar topological feature, causing the rather large intrinsic AHE. Theoretical calculation will give some hints on it. Furthermore, the strong interlayer frustrating coupling provides some new ingredients. For example, the synergic effects of real-space spin chirality and momentum-space Berry curvature give both large THE and AHE in YMn$_{6}$Sn$_{6}$. The former will also have some important influences on the topology of band structure which might be sensitive to the spin texture.\cite{Kuroda,Suzuki}


%
%

In summary, centrosymmetric YMn$_{6}$Sn$_{6}$ with Mn kagome lattice shows a giant THE which may be intimately related to the topological magnetic excitations (SkL) stabilized under field.
Moreover, YMn$_{6}$Sn$_{6}$ also exhibits relative large intrinsic AHE and anisotropic carrier type under different field directions.
Our findings clearly indicates that kagome magnetic metals show the unique topological feature not only in momentum space but also in real space.
More importantly, the rich tunability of various degrees of freedom in this big family of materials provides an excellent opportunity to study the strong entanglement between
frustrated magnetism, electronic correlation and topological electronic structure via spin-orbital coupling.

\begin{methods}

\noindent\textbf{Single crystal growth and structural characterization.} YMn$_{6}$Sn$_{6}$ single crystals were grown by flux method.
X-ray diffraction (XRD) patterns were performed using a Bruker D8 X-ray machine with Cu $K_{\alpha}$ radiation ($\lambda=$ 0.15418 nm).

\noindent\textbf{Magnetization and Transport measurements.} Magnetization and electrical transport measurements were performed in Quantum Design PPMS-14 T. Both longitudinal and Hall electrical resistivity were measured simultaneously in a standard five-probe configuration. For measuring $\rho_{xx}$ [$I\Vert V\Vert (100)$, $B\Vert (001)$] and $\rho_{yx}$ [$I\Vert (100)$, $V\Vert (120)$, $B\Vert (001)$], the crystal with size about 2.0$\times$1.5$\times$0.2 mm$^{3}$ was used and the Pt wires are stick on the sample directly using Ag epoxy. For measuring $\rho_{zz}$ [$I\Vert V\Vert (001)$, $B\Vert (001)$] and $\rho_{yz}$ [$I\Vert (001)$, $V\Vert (120)$, $B\Vert (100)$], the samples were fabricated into rectangular shape using the focus ion beam (FIB) technique by using JEOL JEM-9320FIB.
The W contacts were deposited between the sample and Au pads and then the Pt wires were connected on the Au pads using Ag paste.
In order to effectively get rid of the longitudinal resistivity contribution due to voltage probe misalignment, we measured the Hall resistivity at positive and negative fields. The total Hall resistivity was determined by $\rho_{\rm H}(B)=[\rho_{\rm H}(+B)-\rho_{\rm H}(-B)]/2$. Correspondingly, the longitudinal resistivity was obtained by $\rho(B)=[\rho(+B)+\rho(-B)]/2$.

%
%
%

\end{methods}

\begin{addendum}

\item This work was supported by the National Key R\&D Program of China (Grants No. 2016YFA0300504), the National Natural Science Foundation of China (No. 11574394, 11774423, 11822412), the Fundamental Research Funds for the Central Universities, and the Research Funds of Renmin University of China (RUC) (15XNLQ07, 18XNLG14, 19XNLG17).

\item[Author contributions] H.C.L. provided strategy and advice for the research. Q.W. and Q.W.Y performed the crystal growth, transport and magnetization measurements. Q.W. performed fundamental data analysis; S.F. and H.H. performed the sample fabrication using the FIB technique. H.C.L. wrote the manuscript based on discussion with all the authors.

\item[Supplementary Information] accompanies this paper.

\item[Author Information] The authors declare no competing financial interests. The data that support the findings of this study are available from the corresponding
author H.C.L. (hlei@ruc.edu.cn) upon reasonable request.

\end{addendum}

\begin{figure}
  \centerline{\epsfig{figure=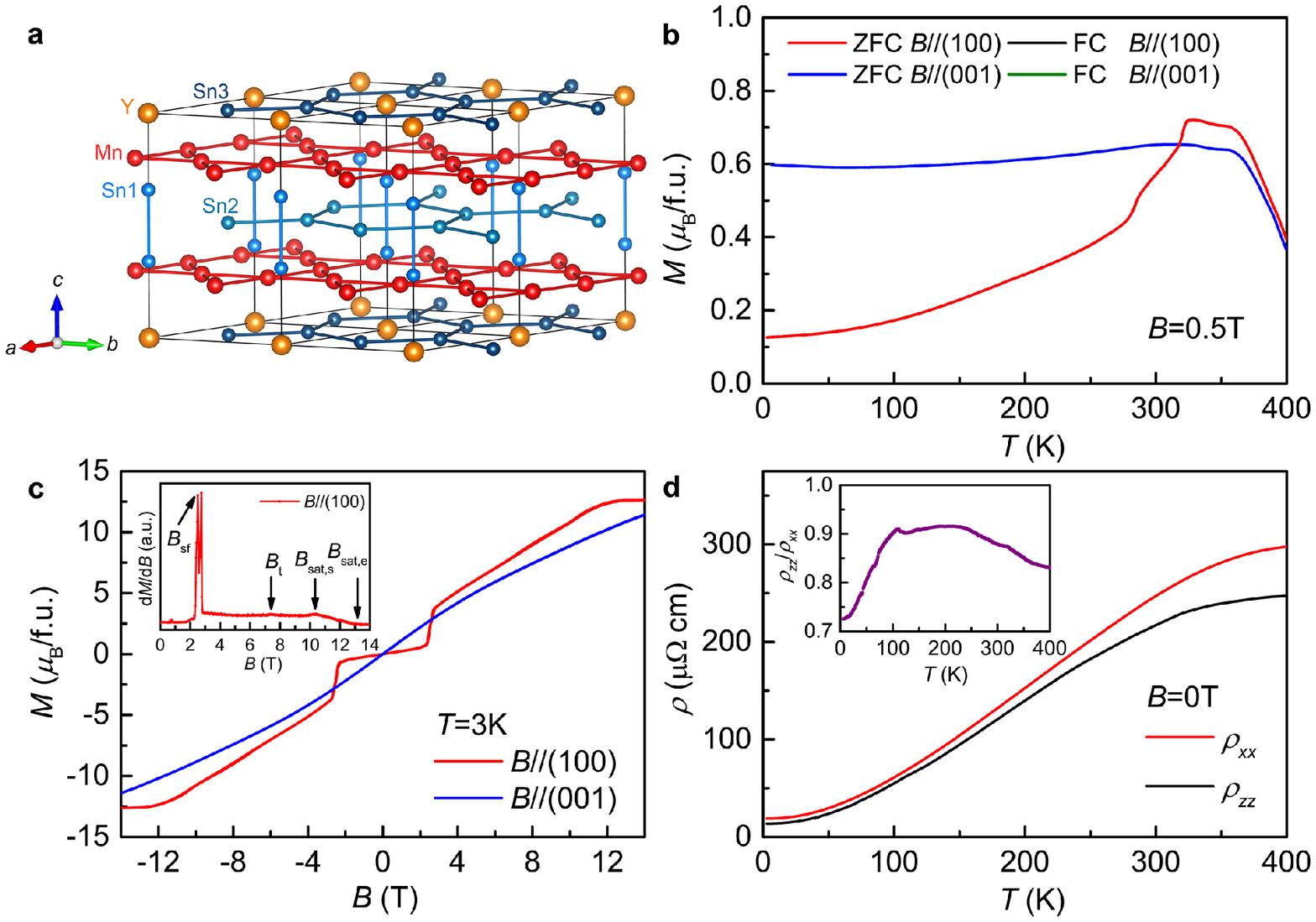,width=0.8\columnwidth}}
  \caption{\textbf{Structure, anisotropic magnetism and longitudinal resistivity of YMn$_{6}$Sn$_{6}$.}
  \textbf{a}, Crystal structure of YMn$_{6}$Sn$_{6}$. The large yellow, medium red and small blue (light blue and dark blue) ball represents Y, Mn and Sn1 (Sn2 and Sn3) atoms, respectively.
  \textbf{b}, Temperature dependence of magnetization $M(T)$ with ZFC and FC modes at $B=$ 0.5 T for $B\Vert (100)$ and $B\Vert (001)$.
  \textbf{c}, Field dependence of magnetization $M(B)$ at $T=$ 3 K for $B\Vert (100)$ and $B\Vert (001)$. Inset: d$M(B)$/d$B$ vs. $B$ for $B\Vert (100)$.
  \textbf{d}, Zero-field longitudinal resistivity $\rho_{xx}(T)$ and $\rho_{zz}(T)$ as a function of $T$. Inset: temperature dependence of the ratio $\rho_{zz}(T)/\rho_{xx}(T)$.}
\end{figure}

\newpage

\begin{figure}
  \centerline{\epsfig{figure=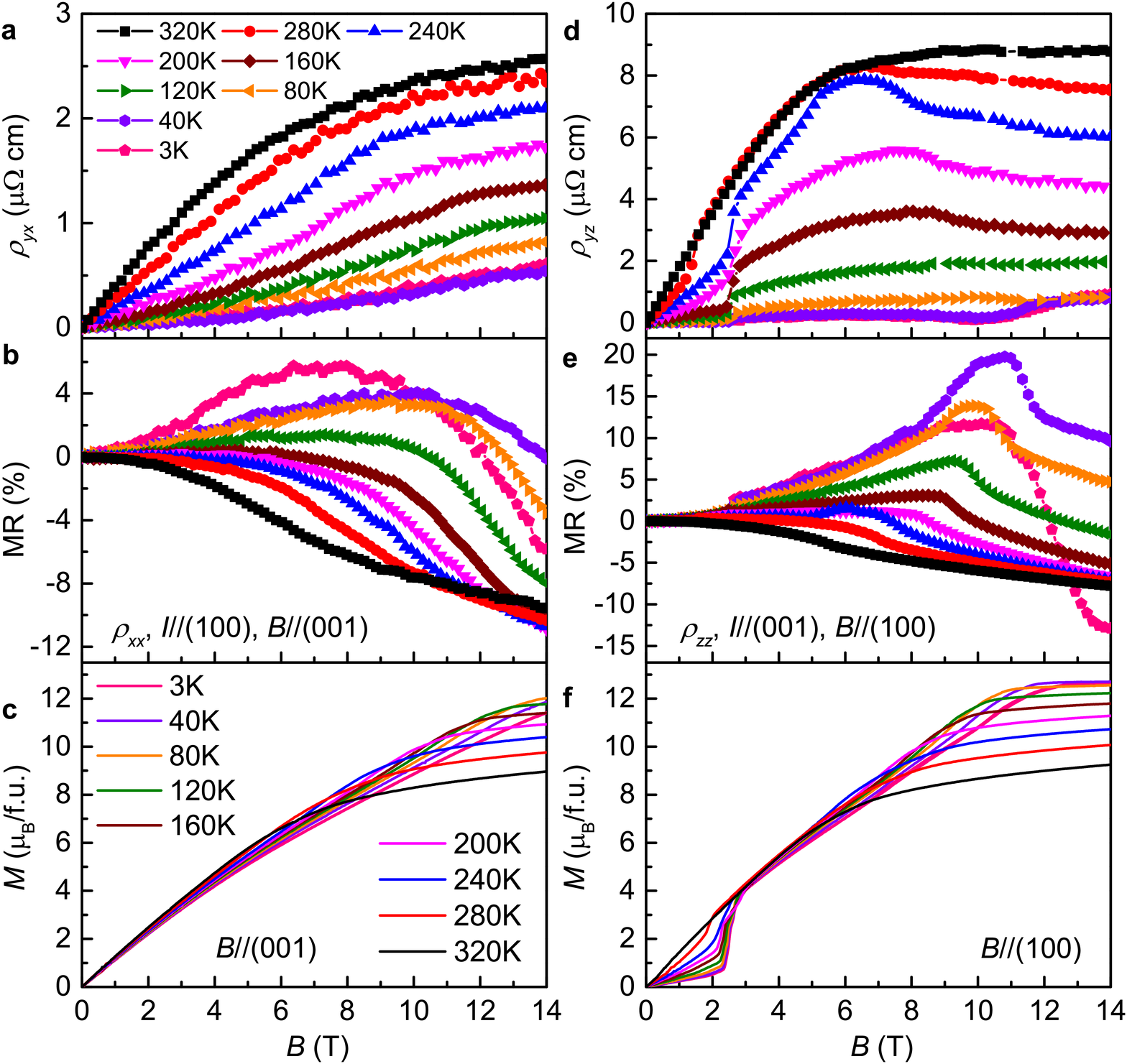,width=0.8\columnwidth}}
  \caption{\textbf{Field dependence of transport and magnetization properties of YMn$_{6}$Sn$_{6}$.}
  \textbf{a}, Hall resistivity $\rho_{yx}(B)$,
  \textbf{b}, Magnetoresistance MR ($= 100\%*[\rho_{xx}-\rho_{xx}(0)]/\rho_{xx}(0)$) and
  \textbf{c}, $M(B)$ as a function of magnetic field at various temperatures when $B\Vert (001)$.
  \textbf{d}, Hall resistivity $\rho_{yz}(B)$, \textbf{e}, Magnetoresistance MR ($= 100\%*[\rho_{zz}-\rho_{zz}(0)]/\rho_{zz}(0)$) and
  \textbf{f}, $M(B)$ as a function of magnetic field at various temperatures when $B\Vert (100)$. The symbols in \textbf{a}, \textbf{b}, \textbf{d}, and \textbf{e} have same color codes. The colors of curves in \textbf{c} and \textbf{f} have same definitions.}
\end{figure}

\newpage

\begin{figure}
  \centerline{\epsfig{figure=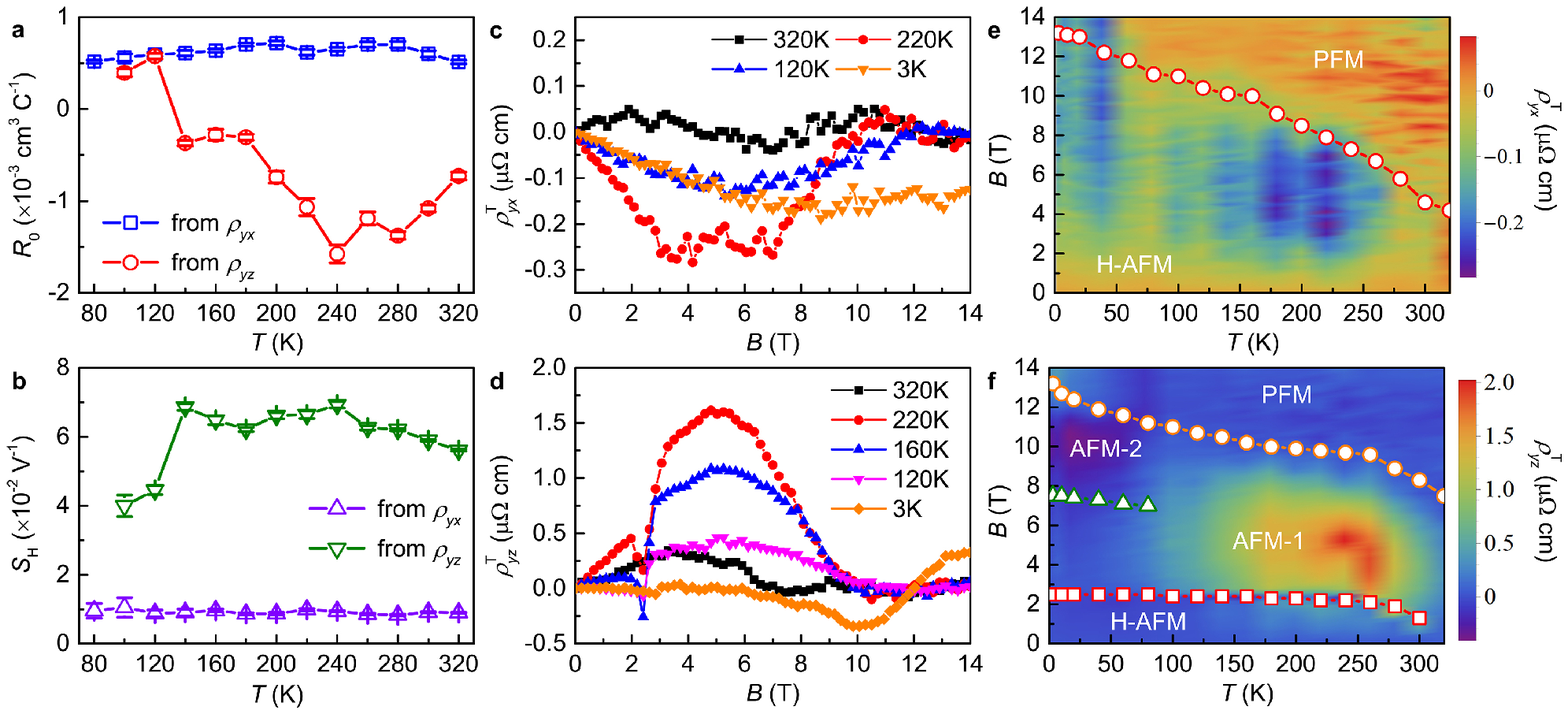,width=0.9\columnwidth}}
  \caption{\textbf{AHE and THE of YMn$_{6}$Sn$_{6}$.}
  Temperature dependence of
  \textbf{a}, $R_{0}(T)$ and
  \textbf{b}, $S_{H}(T)$ derived from $\rho_{yx}(B)$ and $\rho_{yz}(B)$.
  \textbf{c}, $\rho_{yx}^{\rm T}(B)$ and
  \textbf{d}, $\rho_{yz}^{\rm T}(B)$ as a function of $B$ at selected temperatures.
  The contour plot of
  \textbf{e}, $\rho_{yx}^{\rm T}(T,B)$ and
  \textbf{f}, $\rho_{yz}^{\rm T}(T,B)$ for $T$ between 3 K and 320 K and $B$ from 0 to 14 T.
  The red circular symbols in \textbf{e} represent the $B_{\rm sat,s}$. The orange circular, green triangular and red square symbols in \textbf{f} represent the $B_{\rm sat,e}$, $B_{t}$ and $B_{\rm sf}$, respectively. The "H-AFM", "AFM-1", "AFM-2" and "PFM" represents the helical AFM state, the first non-collinear AFM state, the second non-collinear AFM state and the polarised FM state, respectively.}
\end{figure}

\end{document}


\captionsetup[figure]{labelfont={bf},name={Supplementary Figure},labelsep=period}

\section*{Supplementary Figures}


\begin{figure}[thb]
\centerline{\includegraphics[width=1\columnwidth]{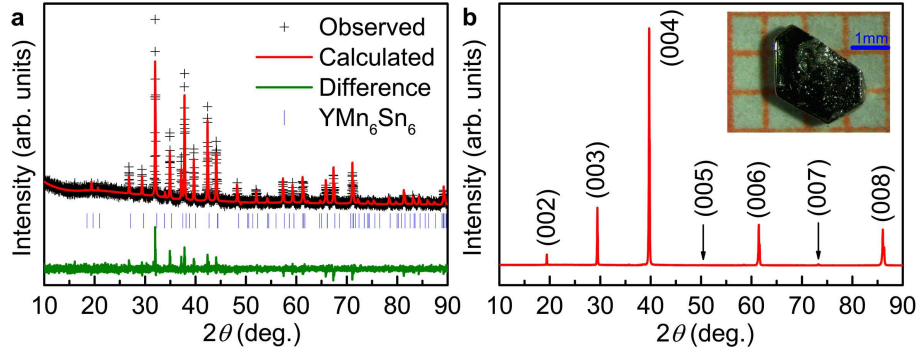}} \vspace*{-0.3cm}
  \caption{\textbf{a}, powder XRD pattern of ground YMn$_{6}$Sn$_{6}$ single crystals. \textbf{b}, XRD pattern of a YMn$_{6}$Sn$_{6}$ single crystal. All of peaks can be indexed by the indices of (00l) lattice planes. It reveals that the crystal surface is normal to the $c$ xaxis with the plate-shaped surface parallel to the $ab$ plane. Inset: photo of typical YMn$_{6}$Sn$_{6}$ single crystals on a 1 mm grid paper.}
\end{figure}


\begin{figure}[htb]
\centerline{\includegraphics[width=1\columnwidth]{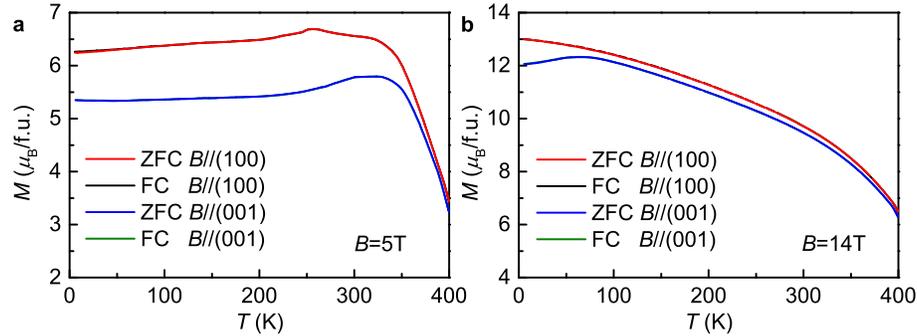}} \vspace*{-0.3cm}
  \caption{Temperature dependence of magnetization $M(T)$ with ZFC and FC modes for $B\Vert(100)$ and $B\Vert(001)$ at \textbf{a}, $B=$ 5 T and \textbf{b}, $B=$ 14 T.}
\end{figure}


\begin{figure}[htb]
\centerline{\includegraphics[width=1\columnwidth]{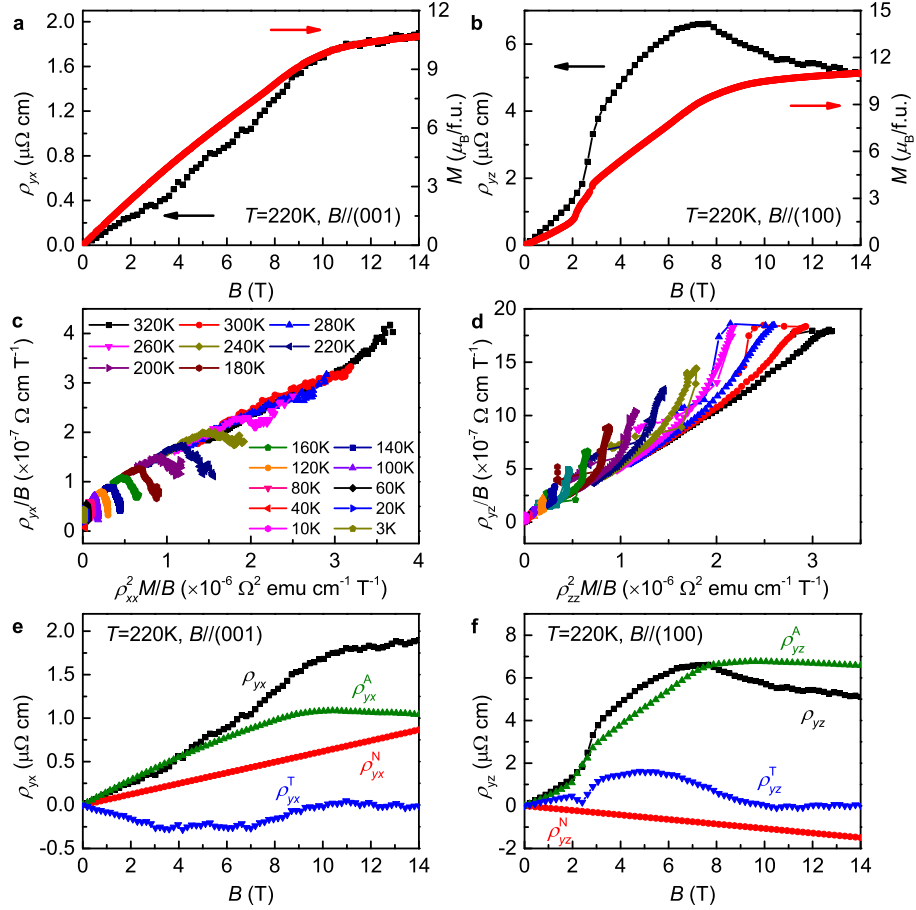}} \vspace*{-0.3cm}
  \caption{Field dependence of \textbf{a}, Hall resistivity $\rho_{yx}$ and $M(B)$ for $B\Vert(001)$ and \textbf{b}, $\rho_{yz}$ and $M(B)$ for $B\Vert(100)$ at 220 K. \textbf{c}, $\rho_{yx}/B$ vs. $\rho_{xx}^{2}M/B$ and \textbf{d}, $\rho_{yz}/B$ vs. $\rho_{zz}^{2}M/B$ at various temperatures. Estimated different components of Hall resistivity for \textbf{e}, $\rho_{yx}$ and \textbf{f}, $\rho_{yz}$. Curves labeled as $\rho_{yx}$ and $\rho_{yz}$ are the total Hall resistivity observed at 220 K. $\rho_{yx}^{\rm N}$ and $\rho_{yz}^{\rm N}$ are the normal Hall resistivity estimated from $R_{0}B$. $\rho_{yx}^{\rm A}$ and $\rho_{yz}^{\rm A}$ are the anomalous Hall resistivity estimated from $S_{\rm H}\rho_{xx}^{2}M$ and $S_{H}\rho_{zz}^{2}M$. $\rho_{yx}^{\rm T}$ and $\rho_{yz}^{\rm T}$ are the topological Hall resistivity derived from $\rho_{yx}^{\rm T}\equiv \rho_{yx} - \rho_{yx}^{\rm N} - \rho_{yx}^{A}$ and $\rho_{yz}^{\rm T}\equiv \rho_{yz} - \rho_{yz}^{\rm N} - \rho_{yz}^{\rm A}$, arising from the spin chirality.  }
\end{figure}


\begin{figure}[htb]
\centerline{\includegraphics[width=1\columnwidth]{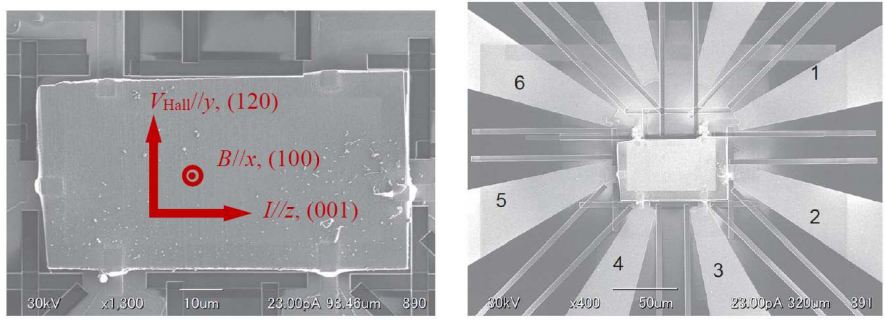}} \vspace*{-0.3cm}
  \caption{The five-probe configuration of $\rho_{yz}$ and $\rho_{zz}$ measurement for the sample fabricated by the focus ion beam (FIB) technique. The contacts 2 and 5 are for $I^{+}$ and $I^{-}$, the contacts 1 and 6 are for $V^{+}$ and $V^{-}$ of $\rho_{zz}$ and the contact 4 and 6 are for $V_{\rm Hall}^{+}$ and $V_{\rm Hall}^{-}$ of $\rho_{yz}$.
  }
\end{figure}


\begin{figure}[htb]
\centerline{\includegraphics[width=0.8\columnwidth]{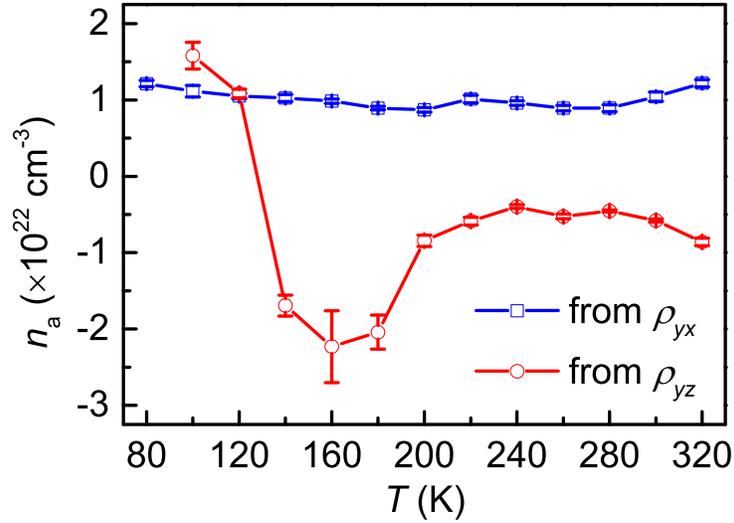}} \vspace*{-0.3cm}
  \caption{Temperature dependence of apparent carrier density $n_{a}$ calculated using the formula $n_{a} = 1/|e|R_{0}$. The $R_{0}$ is derived from $\rho_{yx}$ and $\rho_{yz}$, respectively.}
\end{figure}


\begin{figure}[htb]
\centerline{\includegraphics[width=0.8\columnwidth]{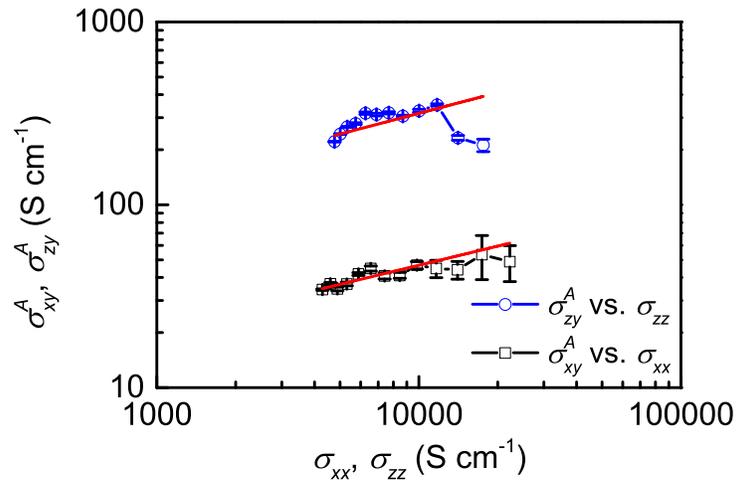}} \vspace*{-0.3cm}
  \caption{Relationship between the anomalous Hall conductivity $\sigma_{\rm H}^{\rm A}$ ($\sigma_{xy}$ or $\sigma_{zy}$) and the longitudinal conductivity $\sigma$ ($\sigma_{xx}$ or $\sigma_{zz}$). The red solid lines represents the fits using the formula $\sigma_{\rm H}^{\rm A}\propto \sigma^{\alpha}$.}
\end{figure}